\documentclass{article}
\usepackage{spconf,amsmath,graphicx}

\def \im [#1]{\textcolor{red}{IM: #1}}
\def \am [#1]{\textcolor{blue}{AM: #1}}
\usepackage{spconf,amsmath,graphicx,booktabs, color, enumitem}
\usepackage{comment}
\usepackage{multirow}

\title{Training sound event detection with soft labels \\  from crowdsourced annotations}
\name{ Irene Mart\'{\i}n-Morat\'o, Manu Harju, Paul Ahokas, Annamaria Mesaros\thanks{This work was supported by Academy of Finland grant 332063 ``Teaching machines to listen". The authors wish to thank CSC-IT Centre of Science Ltd., Finland, for providing computational resources.}}
\address{Computing Sciences, Tampere University,
Tampere, Finland}
\begin{document}

\maketitle

\begin{abstract}

In this paper, we study the use of soft labels to train a system for sound event detection (SED). 
Soft labels can result from annotations which account for human uncertainty about categories, or emerge as a natural representation of multiple opinions in annotation. 
Converting annotations to hard labels results in unambiguous categories for training, at the cost of losing the details about the labels distribution. This work investigates how soft labels can be used, and what benefits they bring in training a SED system. The results show that the system is capable of learning information about the activity of the sounds which is reflected in the soft labels  and is able to detect sounds that are missed in the typical binary target training setup. We also release a new dataset  produced through crowdsourcing, containing temporally strong labels for sound events in real-life recordings, with both soft and hard labels. 
\end{abstract}
\begin{keywords}
Sound event detection, soft labels, crowdsourcing annotations 
\end{keywords}

\section{Introduction}
\label{sec:intro}

Sound event detection has long relied on strongly annotated data for training and evaluation \cite{Mesaros2017,Mesaros2019,serizel2020}, namely data that contains class labels, and temporal boundaries in terms of start time and end time for each event instance. Development of robust methods for SED was hindered by the small size of strongly-labeled datasets \cite{Mesaros2016_EUSIPCO}, motivating the use of alternative training methods based on weak labels (without temporal information) \cite{Dinkel2021}, synthetic data \cite{salamon2014}, or combinations \cite{Turpault2020a}. Currently, the largest strongly-labeled dataset contains 67k clips of 10~s, with an average number of 3.5 sound events per file \cite{Hershey2021_ICASSP}. The length of the clips is artificially small and therefore might misrepresent the true sound events distribution within a longer timeframe. The only datasets with minutes-long recordings are the TUT SED 2016 and 2017 datasets, created by manually annotating all audible event instances and postprocessing of the labels \cite{ Mesaros2017, Mesaros2016_EUSIPCO}.

An often used alternative to manual annotation is crowdsourcing, which was used to produce many datasets for different domains, such as: image classification with ImageNet \cite{Deng2009} or Microsoft COCO \cite{Lin2014}; music classification with OpenMIC \cite{Humphrey2018} or TROMPA\_MER \cite{GomezCanon2022}; audio captioning with CLOTHO \cite{Drossos2020} or audio tagging with MATS \cite{martin2021_EUSIPCO}. 
To benefit from the simplicity of collecting weak labels through crowdsourcing, in our earlier work we proposed a method to estimate temporally strong labels from weakly-labeled overlapping segments of an audio clip \cite{martin2021_WASPAA}. Strong labels are estimated by reconstructing the temporal order of the annotated segments and counting activity indicators for each annotated sound instance.
The obtained count-based activity indicators are values between 0 and 1, representing the aggregated opinion of multiple annotators, which can also be interpreted as reflecting the uncertainty of the annotators pool about the label. These non-binary values can be seen as \emph{soft labels} \cite{Mendez2022}, as opposed to  the normal case of \emph{hard labels} where each event class has a binary activity level associated indicating if the sound is active (1) or not (0).

Human uncertainty represented as soft labels was shown to make classification more robust and improve generalization in image classification \cite{Peterson2019}, helping the model define transitions between ambiguous classes \cite{Grossmann2022}. Soft labels are typically used through labels smoothing regularization \cite{Szegedy2016}, and are also produced in mixup augmentation by interpolating the one-hot encoding training labels \cite{Zhang2018}, leading to a smoother estimate of uncertainty. In contrast to the one-hot classification case, the special characteristic of sound event detection is the multi-label setup, where one time-segment of audio can have one or multiple events present at once. The predicted outputs in the network do not need to sum up to one, which makes the use of soft labels a readily suitable alternative to the usual multi-hot encoding training scenario. 

In this paper we investigate if the soft labels contain information  useful for the training process, and how such information can be used to improve detection performance. The paper presents two main contributions: first, we release a new dataset, MAESTRO-Real, containing temporally strong labels for sound events in real-life environments. The data was annotated using the procedure in \cite{martin2021_WASPAA} and is provided with soft and hard labels. Second, we show that a SED system trained using soft labels is capable of learning the soft input representation, and able to detect under-represented classes that in a normal setup based on hard labels are not detected.

\vspace{-10pt}
\section{Soft labels dataset creation}
\label{sec:method}
Crowdsourcing for annotating sound events in long audio recordings has been proposed in \cite{martin2021_WASPAA} and shown to produce reasonable annotations. The method consists of a single-pass multi-label annotation \cite{Cartwright2019} with weak labels; the presence of a particular sound is explicitly indicated by annotators by selecting the corresponding label from a predefined set of target sounds, while the absence is indicated implicitly by not selecting a label.

\subsection{Annotation and label estimation procedure}
\begin{figure}[!t]
    \centering
    \includegraphics[width=0.9\columnwidth]{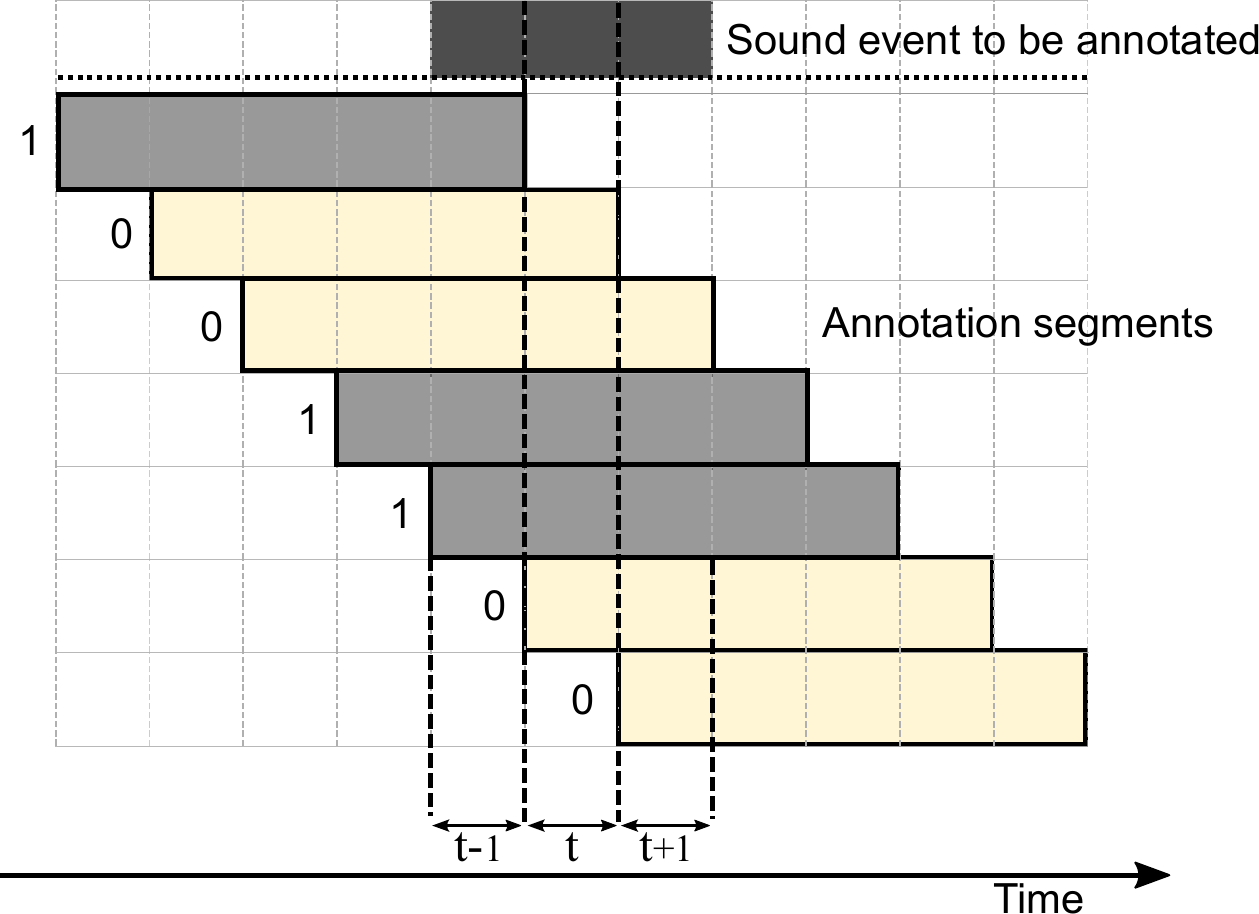}
    \caption{Annotation procedure and event activity estimation: the soft label in segment $t$ takes into account the individual opinions available for the segment, weighed by annotator's competence.}
    \label{fig:ann}
    \vspace{-12pt}
\end{figure}

The annotation process uses weak labeling of densely overlapping segments, and reconstructs the temporal activity of the target sound events as a count-based value based on the annotator opinions. The process is illustrated in Fig. \ref{fig:ann}. The annotation process creates segments $t$ with a length determined by the annotation hop, each such segment being part of multiple annotation segments. For each annotation segment there are multiple opinions. For aggregation, one can use for example majority vote among the $M$ available opinions for segment $t$, to obtain either 0 (sound event inactive) or 1 (sound event active). A one-second annotation hop results in a temporally strong annotation with a resolution of 1~s. While not exactly fitting the strict definition of strong labels, the additional temporal information available in comparison with a weak label (e.g. per 10~s) was shown to bring significant improvement in performance for example to audio event classification \cite{Hershey2021_ICASSP}. 

We aggregate the multiple opinions by taking into account the annotator competence estimated using MACE (multi-annotator competence estimation) \cite{hovy2013} as described in \cite{martin2021_WASPAA}. We use a slightly different procedure than \cite{martin2021_WASPAA} to determine the soft label: instead of using MACE to estimate the ``true" weak label, we use it only for competence estimation. We then process the annotator opinions for each segment $t$ by combining them based on annotators' competence as follows: 

\begin{equation}
    a_t = \frac{{\sum_{j=1}^{M} \left ( \theta_j \cdot v_j \right ) }}{\sum_{j=1}^{M} \theta_j}
    \label{eq:estim}
\end{equation} 

\noindent where $a_{t}$ is the activity level $a$ for one class in segment $t$, $M$ is the number of annotators for that segment, $\theta_j$ is the competence of annotator $j$ estimated using MACE, and $v_j$ indicates the annotator's opinion, being $1$ for the presence and $0$ for the absence of the label. The estimation is done independently for each class. For complete details about the annotation process and the annotator competence estimation procedure, we refer the reader to \cite{martin2023}. 

The resulting value $a_{t}$ is a number between 0 and 1 that can be considered a \emph{soft label}. In this respect, the soft labels are activity indicators for each event class in consecutive segments, which reflect the aggregated uncertainty of the annotators. To produce \emph{hard labels} (binary activity indicators, as typically used in SED training) we apply a 0.5 threshold, i.e. events with activity indicators above 0.5 are considered active (1), while those below 0.5 are considered inactive (0).

The formulation in eq.~\ref{eq:estim} means that a minority of annotators (e.g.~2 of 5) can produce a soft label over 0.5 if they are highly reliable annotators, therefore producing a hard label of 1, even though it is not a majority. This is the intended effect of estimating the reliability of the annotators, and trusting more the more diligent ones.

\subsection{MAESTRO-Real Dataset}
\label{sec:dataset}

\begin{table*}[]
    \centering
    \begin{tabular}{l|l l l l l l l l l l l l l l l l l |l l}
    \toprule
          Scene & 
          \rotatebox{90}{announcement} &\rotatebox{90}{birds singing} & \rotatebox{90}{brakes squeaking} & \rotatebox{90}{car} &\rotatebox{90}{cash register} & \rotatebox{90}{children voices} & \rotatebox{90}{coffee machine} & \rotatebox{90}{cutlery/dishes} & \rotatebox{90}{door opens/closes} & \rotatebox{90}{footsteps} & \rotatebox{90}{furniture dragging} & \rotatebox{90}{large vehicle} &\rotatebox{90}{metro approaching} &\rotatebox{90}{metro leaving} & \rotatebox{90}{people talking}  &
          \rotatebox{90}{shopping cart} &\rotatebox{90}{wind blowing} & \rotatebox{90}{No. of files} & \rotatebox{90}{Duration} \\
         \midrule
         Cafe/restaurant & - & - & - & - & - & 27 & 8 & 71 & - & 6 & 3 & - & - & - & 48 & - & - & 15 & 56.91 \\
         City center  & - & - & 37 & 92 & - & 4 & - & - & - & 62 & - & 117 & - & - & 49 & - & - & 15 & 61.47 \\
         Grocery store  & 0 & - & - & - & 0 & 6 & - & - & - & 27 & - & - & - & - & 61 & 5 & - & 14 & 49.53 \\
         Metro station  & - & - & - & - & - & 9 & - & - & 4 & 116 & - & - & 92 & 99 & 129 & - & - & 14 & 59.21 \\
         Residential area  & - & 60 & - & 76 & - & 10 & - & - & - & 26 & - & - & - & - & 20 & - & 38 & 17 & 59.82 \\
         \midrule
         Total  & - & 60 & 37 & 168 & - & 56 & 8 & 71 & 4 & 237 & 3 & 117 & 92 & 99 & 307 & 5 & 38 & 75 & 286.94 \\
         
         \bottomrule
    \end{tabular}
    \caption{Dataset statistics: event instances according to the hard labels, number and total duration of audio files in minutes.}
        \label{tab:dataset_stats}
        \end{table*}

We produced the MAESTRO-Real (Multi-Annotator Estimated STROng labels) dataset consisting of real-life recordings belonging to five different acoustic scenes: cafe/restaurant, city center, grocery store, metro station and residential area, a subset of TUT Acoustic Scenes 2016 dataset \cite{Mesaros2016_EUSIPCO}. The audio was recorded as 3-5 minutes long clips, which were split into short segments for the release of TUT Acoustic Scenes 2016. In this work we used the original long recordings. 

Each file was annotated using Amazon Mechanical Turk, according to the procedure proposed in \cite{martin2021_WASPAA}. The annotators had to select which sounds were active in a 10~s segment from a list of six classes for each acoustic scene; some event classes are common to all the scenes (e.g. children voices, footsteps, people talking), while others are scene-specific (e.g. birds singing or metro leaving). The collected weak labels were aggregated into temporally strong labels according to eq.~\ref{eq:estim}, resulting in soft labels; these were then converted to hard labels using a threshold of 0.5.

The statistics of the event classes for the different acoustic scenes are presented in Table~\ref{tab:dataset_stats}, based on the hard labels. From the table we notice that the data is highly unbalanced: some classes have a very low number of event instances, while for some classes no instances exist in the hard labels (e.g. cash register which was given as a label for the grocery store files). An inspection of the soft labels reveals that for such classes there are many non-zero values, but they are below 0.5. We speculate that learning from soft labels will take advantage of this additional information and facilitate learning of the under-represented classes to some extent. 

The dataset is publicly available\footnote{MAESTRO-Real, 10.5281/zenodo.7244360} and consists of 75 audio files and their corresponding annotations in two versions: one set of annotations with hard labels and one with soft labels. The hard labels format contains start time, end time, and textual label for each sound instance, with the particular feature that the timestamps are rounded to seconds due to the annotation procedure (e.g. 2 11 car). The soft labels format contains consecutive 1~s segments in a similar format, with start time, end time, and textual label, and in addition a value indicating the soft label for each event class in the given segment (e.g. 2 3 car 0.9; 2 3 footsteps 0.7; 3 4 car 0.85; etc).

\section{Training SED with soft labels}
\label{sec:sed_training}

We employ a basic CRNN architecture that we train in different setups, using the hard or the soft labels for comparison. We used  the same 5-fold training procedure for all the training setups. %

\subsection{SED model}

The model follows the CRNN architecture from \cite{Adavanne2017}, which consists of three convolutional layers and two bi-directional gated recurrent units (GRU), with a sigmoid layer as output layer. As input the model uses mel energies calculated using a window size of 2048 samples with a hop length equivalent to 20 ms, and 64 mel filter banks, with the lower and upper frequencies set to 50 and 14kHz. 
When training with hard labels the system uses BCE loss, which allows independent outputs between 0 and 1. To train with soft labels, we first employ the same BCE loss, forcing the system to predict outputs as close as possible to binary activity indicators, as would be the case when using data augmentation or label smoothing. In addition, we investigate training using MSE loss, to teach the system to predict outputs as close as possible to the provided soft activity indicators instead of binary. In this case, the output layer is linear, to allow for linear regression. 
In the remainder of the paper these setups are denoted as:

\begin{itemize}[noitemsep,nolistsep,leftmargin=*]
    \item $\text{H\_BCE\_sig}$: hard labels for training, BCE loss, and a dense output layer with sigmoid activation.
    \item $\text{S\_BCE\_sig}$: soft labels for training, BCE loss, and a dense output layer with sigmoid activation.
    \item $\text{S\_MSE\_lin}$: soft labels for training, MSE loss, and a dense output layer with linear activation.
\end{itemize}

\begin{figure*}[!t]
    \centering
     \vspace{-10pt}
    \includegraphics[width=1.0\textwidth]{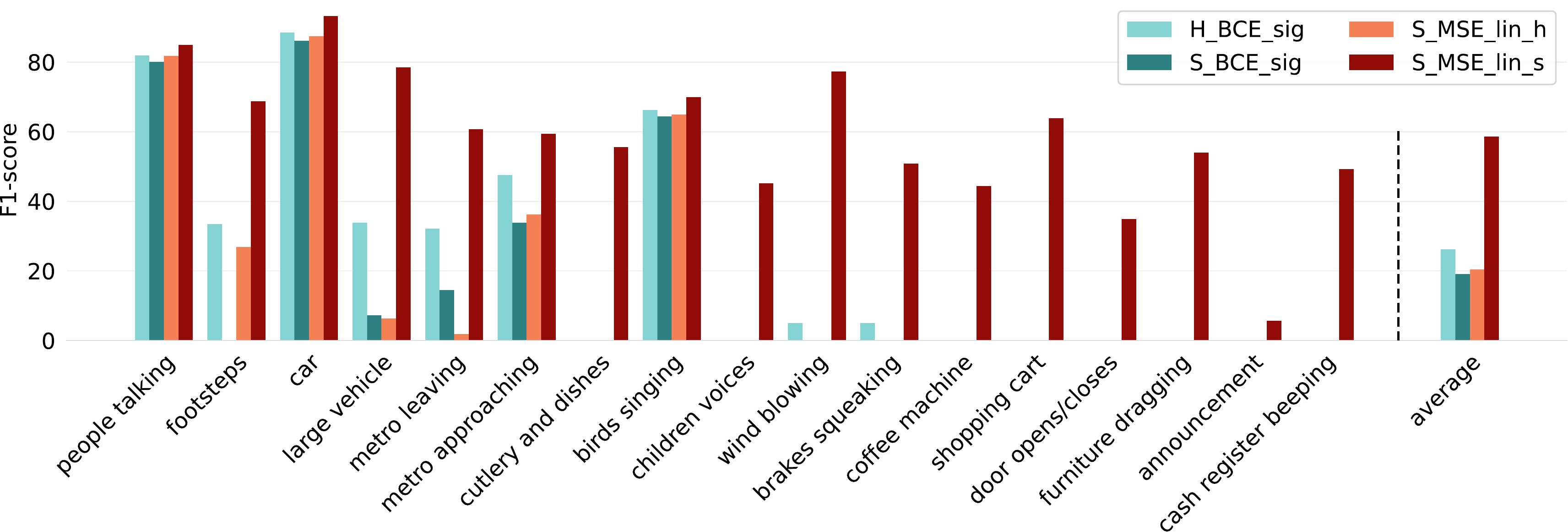}
    \vspace{-20pt}
    \caption{F1-score values for each of the models, with x-axis ordered from higher number of instances to lower number of instances per class. H\_BCE\_sig and S\_MSE\_lin\_s values are calculated using a class-dependent threshold.}
    \label{fig:fscore}
    \vspace{-8pt}
\end{figure*}

\subsection{Evaluation}

We evaluate the outputs by comparing them with the reference labels. The sigmoid outputs are probabilities between 0 and 1, so a threshold of 0.5 is applied on each output to determine the predicted event class activity independently. These outputs are then evaluated against the hard labels using 1~s segment-based F1-score and error rate \cite{Mesaros2019}. 
Because the binary outputs do not explain the uncertainty, we calculate the Kullback-Leibler divergence (KLD) between the soft system output (before threshold) and the reference labels as provided in training. KLD allows a better insight into how much the distribution of the predicted labels differs from the reference annotation \cite{Murphy2012}, and is particularly suitable for comparing the system output to the soft labels. 

\section{Experimental results}
\label{sec:results}

Table \ref{tab:metrics} shows that the performance of the three models is very similar when evaluated using micro-averaged metrics. Based on the values of KLD, we can say that training with BCE and soft labels provides system outputs closer to the reference than training with hard labels, but is detrimental to the overall metrics. Training with MSE and soft labels results in an output distribution close to the target, and highest overall performance, similar to the hard labels case. For the evaluation in Table~\ref{tab:metrics}, the S\_MSE\_lin system outputs were transformed to binary with a 0.5 threshold and compared with the hard labels, but KLD was calculated from the soft outputs to the soft reference labels. 

\begin{table}[]
    \centering
    \begin{tabular}{l|c|cc||c}
    \toprule
     Method    & Threshold & ER & F1  & KLD  \\
     \midrule
     $\text{H\_BCE\_sig}$  & 0.5  & 0.44 & 70.60\% & 3.710  \\
     $\text{S\_BCE\_sig}$  & 0.5  & 0.48 & 68.25\% & 3.334  \\
     $\text{S\_MSE\_lin}$  & 0.5  & 0.44 & 70.82\% & 1.383   \\
     $\text{S\_MSE\_lin}$  & class-dependent  & 0.47 & 67.30\% & --   \\
    \bottomrule
    \end{tabular}
    \caption{Overall ER and F1-score in 1~s segments for the different training setups, evaluated against reference labels. The last column is the KL divergence between the system output (soft value) and the reference (binary for BCE, soft for MSE).}
    \label{tab:metrics}
    \vspace{-10pt}
\end{table}

An in-depth analysis of the class-wise performance shows that the under-represented classes are not detected by the H\_BCE\_sig model. This behavior has been observed in previous work \cite{Adavanne2017}, and is a result of the training process: the system optimizes to learn as well as possible the large classes, and be penalized less by the cross-entropy for ignoring the small classes. The same happens for the H\_BCE\_sig model. 
When trained with MSE, the outputs are expected to be close to the provided soft labels, which in many cases are below 0.5. For this reason, binarizing its output with the 0.5 threshold does not show any improvement. This case is shown in Fig.~\ref{fig:fscore} as the S\_MSE\_lin\_h system, with \textit{h} indicating that reference and system outputs are the hard labels as for the previous two cases.

To understand the dynamics of the output w.r.t. the soft reference labels, we apply a class-wise threshold determined as the trimmed midrange of the soft labels in the training data. This is equivalent to operating point optimization per class in a SED system, except that we apply it on both soft reference and system output. The choice of midrange is motivated by the traditional 0.5 threshold typically used to binarize the sigmoid outputs: 0.5 is the midrange of the reference annotation values (binary activity, 0 or 1), not the mean (which would depend on the actual distribution of the values). Because we are working with human opinions, we use a 10\% trimmed midrange, to make the value more robust to outliers. 

The results are presented in Fig.~\ref{fig:fscore} as the S\_MSE\_lin\_s system, with \textit{s} indicating that we compare the output with soft reference labels. It becomes apparent that the system does detect activity for all the classes, but this output is lost in the postprocessing when using the same threshold for all the classes. Class-wise average F1-score in this case is clearly higher than the others. Micro-average ER and F1-score for this case are shown in Table \ref{tab:metrics}, 0.47 and 67.3\%, respectively, indicating that there are overall more errors (higher ER) as a trade-off to detecting activity for all the classes. 

\section{Conclusions}
\label{sec:concl}
Sound event detection is still one of the most challenging tasks in environmental audio analysis. Because there are few available datasets with long recordings, recent studies focused on using heterogeneous data for training, with evaluation based on small-size, manually annotated data. With this work we contribute to the data for SED by providing a new dataset consisting of five real-life acoustic scenes, with 17 sound event classes annotated through crowdsourcing, including soft and hard labels. We also showed that soft labels can be successfully used to learn acoustic models, and system outputs can be processed to produce hard labels with individual per-class thresholds.
Future work will consider tailoring models to the unbalanced nature of the datasets, which may arise from sound events being genuinely rare, or due to some sounds  being recognized only by a minority of the annotators.


\bibliographystyle{IEEEbib}{}

\bibliography{refs}

\end{document}